\documentclass[letterpaper, 10pt, conference]{ieeeconf} 
\IEEEoverridecommandlockouts 
\usepackage{textcomp} 
\usepackage{arydshln} 
\usepackage{xcolor}
\usepackage{fontenc}
\usepackage{graphicx}
 \usepackage{latexsym}
 \usepackage{amssymb}
 \usepackage{amsmath}
 \usepackage{multirow}
 \usepackage{epsfig}
 \usepackage{epstopdf}

\usepackage{caption}
\usepackage{subcaption}

\usepackage{graphics} 
\usepackage{epsfig} 
\usepackage{tikz}
\usetikzlibrary{arrows}
\usepackage{circuitikz} 
\usepackage{float} 
\usepackage{pgfplots}
\usepackage{pdfpages} 

\usepackage{filecontents}
\usepackage{mathptmx} 
\usepackage{times} 
\usepackage{amsmath} 
\usepackage{amssymb} 

\renewcommand{\baselinestretch}{1.0}

\newtheorem{theo}{\noindent Theorem}[section]

\newtheorem{corollary}[theo]{Corollary}

\newtheorem{example}[theo]{Example}
\newtheorem{lemma}[theo]{Lemma}

\newtheorem{remark}[theo]{Remark}

\newcommand{\scalemath}[2]{\scalebox{#1}{\mbox{\ensuremath{\displaystyle #2}}}} 

\newcommand{\prthm}[1]{{\em \noindent Proof of Theorem \ref{#1}:}} 
 
\newcommand{\EP}{\hfill $\Box$ \vspace{1mm}\newline}

\newcommand{\Lap}{{L}}
\newcommand{\Lapy}{{L^y}}
\newcommand{\Lapx}{{L^x}}
\newcommand{\Lapsig}{{L}_\sigma}
\newcommand{\Lapsigt}{{\tilde{L}}_\sigma}
\newcommand{\Graph}{\mathcal{G}}
\newcommand{\Vrtx}{\mathrm{V}}
\newcommand{\Edge}{\mathcal{E}}
\newcommand{\Nei}{\mathcal{N}}
\newcommand{\Gammasig}{\Gamma_\sigma}
\newcommand{\Gammasigt}{\tilde{\Gamma}_\sigma}
\newcommand{\Binpy}{B^{y}}
\newcommand{\Binpx}{B^{x}}
\newcommand{\Real}{\mathbb{R}}

\title{\LARGE \bf Stability and Equilibrium Analysis of Laneless Traffic with Local Control Laws}

\author{ Rakesh U. Chavan, Debraj Chakraborty, and
  D. Manjunath\\ 
  Department of Electrical Engineering, Indian Institute of Technology
  Bombay. \\
  e-mail: \{rakeshchavan,dc,dmanju\}@ee.iitb.ac.in
  }

\makeatother

\begin{document}

\maketitle
\begin{abstract}
In this paper, a new model for traffic on roads with multiple lanes
is developed, where the vehicles do not adhere to a lane discipline.
Assuming identical vehicles, the dynamics is split along two independent
directions\textemdash{}the $Y$-axis representing the direction of motion
and the $X$-axis representing  the lateral or the direction perpendicular
to the direction of motion. Different influence graphs are used to
model the interaction between the vehicles in these two directions.
The instantaneous accelerations of each car, in both $X$ and $Y$ directions,
are functions of the measurements from the neighbouring cars according
to these influence graphs. The stability and equilibrium spacings
of the car formation is analyzed for usual traffic situations such
as steady flow, obstacles, lane changing and
rogue drivers arbitrarily changing positions inside the formation.
Conditions are derived under which the formation maintains stability and the desired 
intercar spacing for each of these traffic events. Simulations for some of
these scenarios are included.\end{abstract}

\section{Introduction}

\label{sec:intro}

This paper is motivated by the desire to develop models to analyze
traffic when the roads are wide and the lanes are blurred. Such
traffic behavior is not uncommon in many roads in India where, to
misquote J K Galbraith, a \textquoteleft{}functioning
anarchy\textquoteright{} prevails. We develop a stylized model to
describe the local interaction among the vehicles on such roads and
use this microscopic description to characterize the emerging
macroscopic behaviour. In effect, we consider a multilane system
except that we assume that there is no strict demarcation of lanes and
that vehicles are affected by those in a `cone' rather than by the
vehicle right ahead. On such roads, multiple cars drive abreast but do
not adhere to lane discipline. Toward this we adapt the single lane
model of \cite{GerHub75} to develop our multilane model.
Specifically, a directed graph is used to model the influence on the
acceleration of a vehicle by others in its neighbourhood, i.e., those
that are in its cone. We seek an equilibrium analysis of the dynamical
system model that we develop. Our notion of stability refers to the
condition that all cars attain the same velocity as the leader. Since
this is a car following model, we also have the notion of levels and
the analysis will also obtain conditions such that cars in one
\textquoteleft{}level\textquoteright{} maintain a fixed spacing from
cars in the next \textquoteleft{}level.\textquoteright{} Our analysis
will primarily use the Laplacian of the directed graph that models the
influences, which in turn will allow us to dissolve the lanes, so to
say. The influence graph can also be a weighted graph to enable us to
model the relative degrees of the influences.

Models for vehicular systems have been widely studied both from a
microscopic \cite{GaHeRo} \cite{GeHu75} as well as from a macroscopic
\cite{Hel95} \cite{Pay79} perspective. Various forms of stability
such as input-to-state \cite{TaKuPa}, string stability \cite{SwaHed96}
and mesh stability have been considered. Mesh stability ensures damping
of perturbations in vehicle formations, as they move away from the
source \cite{SePaHe}\cite{PaSeHe02}\cite{SoZh12}. Various applications
are studied namely obstacle avoidance in \cite{SaMu03} and lane changing
in \cite{LaDa} \cite{HaOzRe}. In \cite{KlWe} multilane model for
vehicular traffic is considered and a fluid model is developed. Unlike
string or mesh stability theories, where the main concern is magnification
of disturbances down infinite chains of cars, in this paper, we study
finite number of cars. While it is relatively easy to show stability
properties in this case, we are concerned primarily about equilibrium
spacing between the cars. We show that it is possible, with purely
distributed control laws, to achieve and maintain desired spacing,
even in an ad-hoc laneless traffic situation under a variety of common
traffic conditions and disturbances.

The results presented here, as well as the tools used to derive those
results, are heavily dependent on the theory of consensus in multiagent
systems (see e.g \cite{FaMu04} \cite{JaLiMo03}\cite{SaMu04}\cite{bookReBe}
and the references therein). While \cite{RoSaHe}
has studied vehicle consensus, \cite{ReBe05}\cite{Mo05} derived
stability conditions for time varying topologies. Most of our analysis
depend on these recently developed theories and particularly on \cite{RenAtk07},
where arbitrary vehicle formations (not necessarily unidirectional
as in roadways) have been studied. We specialize these results for
ad-hoc laneless road traffic, and in the process derive distributed
control laws that preserve inter-vehicle spacing and stability under
time-varying formations, switched influence graphs and impulse changes
in driving objectives. While these properties are hard to derive for
general types of motion, it becomes relatively simple here due to
typical uni/bi-directional constrained motion possible on roadways.

The rest of the paper is organized as follows. In
Section~\ref{sec:prelim} we set up the notation and our assumptions on
the models for the $X$ and the $Y$ directions. We then describe the
control laws that are used by the vehicles for each of $X$ and $Y$
directions. In Section~\ref{sec:unidirec:motion}, we analyze the
dynamics of motion in the $Y$ direction and obtain a stability
criterion. We also characterize the influence graph. A similar analysis
is carried out for the $X$ direction in Section~\ref{subsec:x:axis:mot:uni}. In
Section~\ref{sec:time:vary:graph} we present the analysis when the
influence graph varies with time. In
Section~\ref{sec:stab:impulse:effect} we analyze the effects of
impulses on the stability of the formation. Finally,
in Section~\ref{sec:simul} we present some simulation results.

\section{Traffic Model and Control Laws}
\label{sec:prelim} 
\label{sec:sys:assump}

As discussed in the introduction, we are interested in analyzing the
motion of a finite number of cars along roads which has no lane
demarcations. In other words, cars can, in principle, occupy arbitrary
positions with respect to other cars as long as they do not collide
with each other or with the side of the road. However, all the cars
want to travel along the road in one direction and would also like to
keep safe a distance from all its neighbours based on visual feedback
about the position and velocity of the neighbouring cars. We name the
direction of travel along the road as the $Y$-axis and the direction
perpendicular to the road as the $X$-axis. Though in reality, a car's
ability to maneuver in these two directions is coupled, for simplicity
we assume that the dynamics of each car along these two directions are
independent. Moreover, the control laws for each of these directions
consider a different set of influencing neighbours.  This later assumption
is realistic since a driver typically looks to the side before moving
sideways, while he considers only the cars roughly in front of him for
normal forward motion. In effect, the influence graphs for the $X$ and
$Y$ directions are different in our calculations.  In addition, we
also assume, again for simplicity, that the cars in the formation are
identical. We state the various modeling assumptions along these two
directions after a brief description of graph theoretic notation.

\subsection{Notation}

\label{sec:graph:prelim}

The sets of naturals, reals, positive reals and real $n$ tuples are
denoted by $\mathbb{N}$, $\Real$, $\Real^{+}$ and $\Real^{n}$.
An undirected \emph{graph} $\mathcal{G}=(\mathcal{V},\mathcal{E},w)$
is a finite set of nodes $\mathcal{V}$ connected by a set of edges
$\mathcal{E}\subset\mathcal{V}\times\mathcal{V}$ along with a function
$w: \Edge \rightarrow \mathbb{R}^+$.
When two nodes
$a_{i}\in\mathcal{V}$ and $a_{j}\in\mathcal{V}$ are connected to
each other the graph $\mathcal{G}$ said to have an edge between $a_{i}$
and $a_{j}$, denoted by $(a_{i},a_{j})\in\mathcal{E}$. A graph is
said to be \emph{connected} when there exists a path between any two
nodes. A \emph{spanning tree} is a connected graph $\mathcal{G}_{tree}=(\mathcal{V}_{tree},\mathcal{E}_{tree})$
having no cycles in the graph. If the edges of a graph $\vec{\mathcal{G}}=(\mathcal{V},\vec{\mathcal{E}})$
are directed i.e. $(a_{i},a_{j})\in\vec{\mathcal{E}}\nRightarrow(a_{j},a_{i})\in\vec{\mathcal{E}}$,
the graph is called a \emph{directed graph }(or \emph{digraph})\emph{.}
A \emph{rooted directed tree }is a digraph such that there exists
a node (called \emph{root}) and a directed path from that node to
all other nodes in the digraph. A digraph $\vec{\mathcal{G}}$ is
said to contain a \emph{directed spanning tree}, if there exists a
rooted directed tree $\vec{\mathcal{G}}_{tree}=(\mathcal{V}_{tree},\vec{\mathcal{E}}_{tree})$
such that $\mathcal{V}_{tree}=\mathcal{V}$ and $\vec{\mathcal{E}}_{tree}\subseteq\vec{\mathcal{E}}$.
  We define the Laplacian ($\Lap$) for a directed graph with weights $w_{ij}$   as follows:     
   $\ell_{ij} := -1 \quad \mbox{if} \quad j \rightarrow i \quad \mbox{and} 
   \quad \ell_{ii} =\sum_{j=1}^{n} w_{ij} := \mbox{indegree}$.

\begin{figure}
\begin{subfigure}[b]{0.25\textwidth} \includegraphics[scale=0.8]{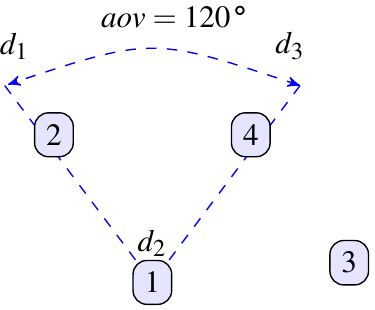}
\caption{$Y$ direction motion: car $3$ is \protect \\
 not seen by car $1$ }

\label{fig:car:view:angle:y} \end{subfigure}\begin{subfigure}[b]{0.25\textwidth}
\includegraphics[scale=0.8]{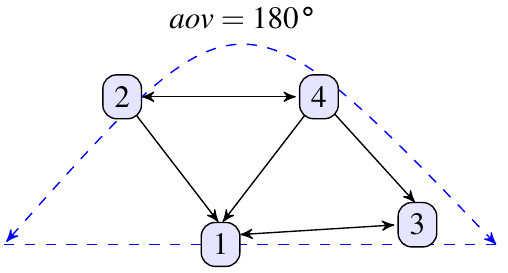} \caption{$X$ direction motion: black lines indicate the underlying graph }

\label{fig:car:view:angle:x} \end{subfigure}\caption{Different influence graphs for $X$ and $Y$ direction depending
on the angle of view ($aov$) of the car}
\vspace{-10pt}
\end{figure}

\subsection{Assumptions for $Y$-direction}\label{sub:Assumptions-for--y}

A1) For determining the influence in $Y$ direction we consider each
driver  to have a fixed viewing angle ($aov$) of $120$\textdegree{} and 
a car must be within this viewing angle to influence the driver. This forms the conical section
$d_{1}d_{2}d_{3}$ as shown in Figure \ref{fig:car:view:angle:y}.
In this figure cars 2 and 4 lie within the conical section $d_{1}d_{2}d_{3}$
formed by car 1 and hence can influence car 1. Car 3 being 
outside the cone cannot influence car 1. 

A2) \label{assump:phantom:car} We assume that the convoy follows
a fictitious leader in the direction of travel. As discussed later,
the role of the leader node is to set a desired velocity for our formation.
This fictitious leader can be considered as a mathematical representation
of velocity limiting rules, regulations or driver experience in practice.
In Figure \ref{fig:graph:info:flow:y:3:cars} node $0$ is the leader.

A3) The influence graph for $Y$ axis is assumed to have a directed
spanning tree with node $0$ as the root. This assumption is equivalent
to connectedness of the influence graph in this framework.

A4) Cars with same number of
hops from the leader node are referred to as being in the same `level'.
In Figure \ref{fig:graph:info:flow:y:3:cars} cars $1,2,$ and $3$
 belong to level one. 

A5) We assume that influence of any car extends up to a maximum of
one level for ease of exposition. This is elaborated on in later sections.

A6) \label{assump:car:num} The convention used for numbering cars
in the graph is as follows: (i) Cars are numbered according to their
levels with cars in higher levels having higher numbers. (ii) Cars
in the same level are numbered from left to right with left most car
being the highest number in that level. 

A7) \label{assump:same:weight:y}We assume that each node can choose the distribution of weights
on its incoming links. This is related to the driver's discretion
about relative importance of the various cars in front of him.

\subsection{Distributed Control Laws for $Y$-direction\label{sec:cntrl:law}}

Consider a convoy of cars represented by influence graphs as shown
in Figure \ref{fig:graph:info:flow:y:3:cars} along $Y$ direction,
where node $0$ is assumed to move at an arbitrary velocity $v_{0}$.
We assume that in the direction of travel, the last car has the lowest
nonnegative $y$ coordinate: car $9$ will have the lowest nonnegative
$y$ coordinate and car $0$ has the highest $y$ coordinate in Figure \ref{fig:graph:info:flow:y:3:cars}. Let
the velocity and position of car $i$ for motion in $Y$-direction
be denoted by $v_{yi}$ and $y_{i}$. The proposed control law model for car $i$
in $Y$ direction is a modified version of single lane car following model of \cite{GerHub75} and is as follows,
\begin{align}\label{equ:ctrl:one:car:y}
\dot{y_{i}} & =v_{yi}\nonumber \\
\dot{v}_{yi} & :=\sum_{j=\Nei(i)}b(w_{ij} v_{yj}- w_{ii}v_{yi})+k(w_{ii}y_{j}-w_{ii}y_{i}-g_{y})
\end{align}
Here $\Nei(i)$ represents the neighbour set of car $i$,
$w_{ij}$ is the weight given to link connecting node $j \to i \ \forall \ j \ne i$, 
$w_{ii} = \sum_{j=1}^n w_{ij} =: W$,
 $b$ and
$k$ are constants, and parameter $g_{y}$ is a tuning parameter used
to adjust the equilibrium distance between cars in consecutive levels.
Consider an $n$ car formation with node $0$ as leader moving along
the road. Let $y\in\Real^{n+1}$ represent the $y$ coordinates or
absolute position of $n+1$ cars. $v_{y}\in\Real^{n+1}$ represent
the velocities of the cars. 

A graph $\Graph=(\Vrtx,\Edge,w)$ having a node set $\Vrtx$ and an edge
set $\Edge$ is used to denote the influence diagram between various cars as per A1.
Each node in the graph represents a vehicle (agent).
The directed edges are introduced as follows: If vehicle $j$ can
be sensed by vehicle $i$ ($i\ne j$), then an edge from $j$ to $i$
exists with weight $w_{ij}$ and is denoted by $j\rightarrow i$. Then the control law in
\eqref{equ:ctrl:one:car:y} can be written as follows: 
\begin{equation}
\begin{bmatrix}\dot{y}\\
\dot{v_{y}}
\end{bmatrix}=\begin{bmatrix}\mathrm{0} & \mathrm{I}\\
-k\ \Lapy & -b\ \Lapy
\end{bmatrix}\begin{bmatrix}{y}\\
{v_{y}}
\end{bmatrix}-kg_{y}\begin{bmatrix}\mathbf{0}\\
\mathbf{1}
\end{bmatrix}\label{equ:control:law:y:direction}
\end{equation}
where $\mathbf{1}\in\mathbb{R}^{n},\ \mathbf{0}\in\mathbb{R}^{n+2},$ $\Lapy$ denotes the Laplacian of the directed
weighted influence graph.

\subsection{Assumptions for $X$-direction}\label{sub:sec:assump:for:x}

B1) For motion in $X$-direction the angle of viewing extends to $180$\textdegree{}.
This is reasonable since a driver would usually look sideways before
making a sideways deviation. In Figure \ref{fig:car:view:angle:x}
car $1$ is influenced by cars $2,3,$ and $4$. 

\begin{figure}
\begin{subfigure}[b]{0.25\textwidth}

\includegraphics[scale=0.75]{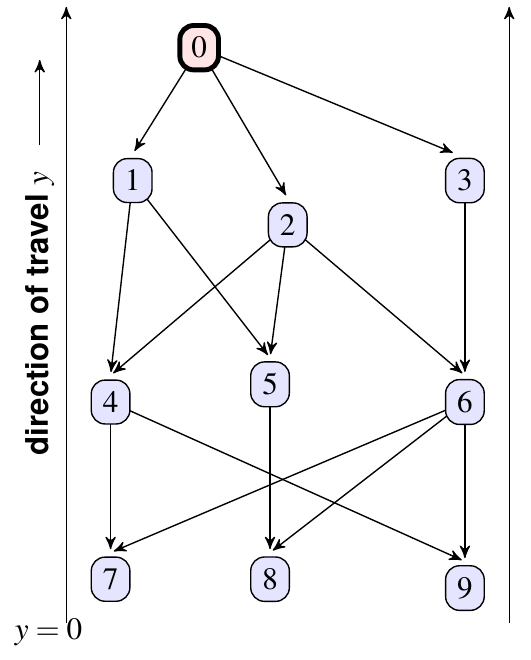} \caption{{Information flow along\\ $Y$ axis } }

\label{fig:graph:info:flow:y:3:cars}

\end{subfigure}\begin{subfigure}[b]{0.25\textwidth} 
\includegraphics[scale=0.7]{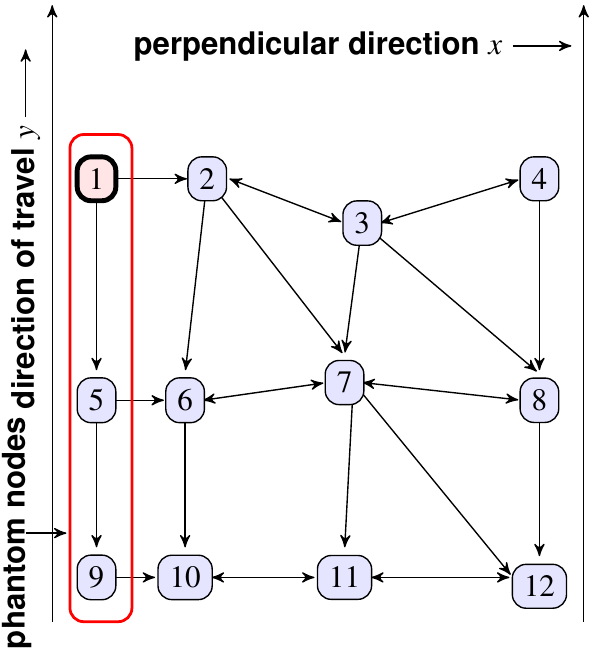} \caption{{Information flow along\\ $X$ axis} }

\label{fig:bidir:cars_multi_lane_x_3_cars}

\end{subfigure}\caption{Figures show the influence graphs for both directions }
\end{figure}

B2) \label{assump:boundar:horiz} For horizontal motion, the boundary
of the road needs to be incorporated into the model. This is achieved
by assuming some fictitious cars are moving along the boundary of the
road at each `level'. Levels are still defined using the $Y$-direction
influence graph and hops from the pseudo $0$-node. In Figure \ref{fig:bidir:cars_multi_lane_x_3_cars}
cars $1,5,9$, enclosed by the red block, represent the boundary of
the road. These nodes have same dynamics as the regular cars but move
under slightly specialized laws: (i) these nodes have no horizontal
velocity (ii) they continue to have the same $y$-coordinate
and the same $y$ velocity as the leftmost car in the corresponding level
(iii) boundary nodes are influenced only by other boundary nodes from
the levels directly above them. 
In Figure \ref{fig:bidir:cars_multi_lane_x_3_cars}
car $5$ is assumed to be influenced  only  by car $1$ and car $9$
only by car $5$. Although the more general case of roads having boundaries
on both sides can be incorporated in the model, here we analyze the
system with boundary on one side of the road for simplicity of exposition.

B3) For $X$ axis motion node $1$ is assumed to play
the role of the leader node as shown in Figure \ref{fig:bidir:cars_multi_lane_x_3_cars}. 
This again sets the $x$ velocity of the formation. 

B4) The influence graph for $X$ direction is assumed to have a directed
spanning tree with node $1$ as root. 

\subsection{Distributed Control Laws for $X$-direction\label{sec:cntrl:law-1}}

Let the velocity and position for car $i$ be denoted by $v_{xi}$
and $x_{i}$. The proposed control law for car $i$ in $X$-direction is%
\begin{align}\label{equ:ctrl:one:car:x}
&\dot{x_{i}}  =v_{xi}\nonumber \\
&\dot{v}{}_{xi}  :=\sum_{j=\Nei(i)}b_x(w_{ij}v_{xj}-w_{ii}v_{x(i)})+k_x(w_{ij}x_{j}-w_{ii}x_{i}-z_ig_{x})
\end{align}%
where $\Nei(i)$ represents the neighbour set of car $i$,
$w_{ij}$ is the weight given to link connecting node $j \to i \ \forall \ j \ne i$,
$w_{ii} = \sum_{j=1}^n w_{ij} =: W$,
$b_x$ and
$k_x$ are constants, and parameter $g_{x}$ is tuned
for desired equilibrium distance between cars in the same level.
This law differs from the $Y$-axis law in that it requires some extra parameters $z_i$'s to
ensure the desired inter-car spacing.
The $X$ axis motion of $n$ cars in the formation is described below.
Let $x\in\Real^{n}$ denotes the coordinates and $v_{x}\in\Real^{n}$
represent the velocities in the $X$ direction. The control law in
\eqref{equ:ctrl:one:car:x} is as follows: 
\begin{equation}
\begin{bmatrix}\dot{x}\\
\dot{v}_{x}
\end{bmatrix}=\begin{bmatrix}\mathit{0} & \mathrm{I}\\
-k_x\ \Lapx & -b_x\ \Lapx
\end{bmatrix}\begin{bmatrix}{x}\\
{v_{x}}
\end{bmatrix}+kg_{x}\begin{bmatrix}\mathbf{0}\\
C
\end{bmatrix}\label{equ:control:law:x:axis}
\end{equation}
where $\mathbf{1}\in\mathbb{R}^{n}$ and $\mathbf{0}\in\mathbb{R}^{n}\mbox{ and }\mathit{0}\in\mathbb{R}^{n\times n},\ 
C\in\Real^{n}$.
\[
C:=\begin{bmatrix}0 & z_{2}\ \hdots\ z_{k} & 0 & z_{k+1}\hdots & 0 & \hdots\ z_{n}\end{bmatrix}
\]
where the nonzero $z_{1},\ z_{2},\hdots\ z_{n}$ are constants locally computed
by each car to ensure a spacing of $g_{x}$ between cars in the same
level. The $\Lapx$ denotes the directed weighted Laplacian of the $X$-axis
influence graph. The dependence of $z_{i}$
 and $g_{x}$ is elaborated on in later sections. 
\begin{figure}
\centering
 \includegraphics[scale=0.8]{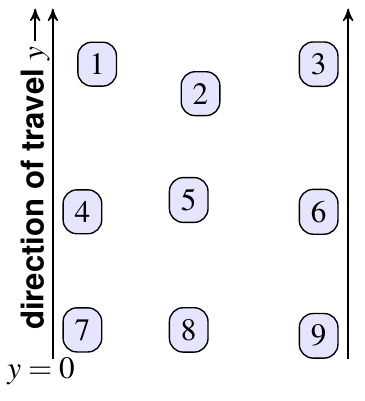} \caption{The convoy of cars 
 for which the influence graphs are 
 represented below}
 \vspace{-10pt}
\end{figure}

\section{Analysis of $Y$-dynamics}

\label{sec:unidirec:motion}\label{subsec:y:axis:mot:uni}

In this section we give conditions under which control law \eqref{equ:control:law:y:direction}
 gives the desired spacing with
unidirectional communication for $Y$  axis motion. We rewrite
\eqref{equ:control:law:y:direction} with the position ($y_{0}(t)$)
and velocity ($v_{0}(t)$) of leader node $0$ as external inputs:
\begin{equation}
\begin{bmatrix}\dot{y}\\
\dot{v_{y}}
\end{bmatrix}=\begin{bmatrix}\mathrm{0} & \mathrm{I}\\
-k\ \tilde{\Lap}^y & -b\ \tilde{\Lap}^y
\end{bmatrix}\begin{bmatrix}{y}\\
{v_{y}}
\end{bmatrix}-kg_{y}\begin{bmatrix}\mathbf{0}\\
\mathbf{1}
\end{bmatrix}+\begin{bmatrix}\mathbf{0}\\
\Binpy
\end{bmatrix}\begin{bmatrix}bv_{0}\\
ky_{0}
\end{bmatrix}\label{equ:control:law:y:direction:leader:inp:main}
\end{equation}
where $\Binpy \in\Real^{n\times2}$, $y,\; v_{y}\in\Real^{n}$. $\tilde{\Lap}^y$
is the reduced Laplacian obtained from $\Lapy$ after removing the
row and column corresponding to node $0$. $\Binpy$ contains the columns
from $\Lapy$ which denote the links between the leader node $0$ and
the remaining cars in the formation. For Figure \ref{fig:graph:info:flow:y:3:cars},
assuming unit weights, $\Binpy$ is given by, 
\begin{align}
\Binpy=\begin{bmatrix}1 & 1 & 1 & 0 & 0\hdots0\\
1 & 1 & 1 & 0 & 0\hdots0
\end{bmatrix}^{T}\label{equ:x:binp:val:exmp-1}
\end{align}
In \eqref{equ:x:binp:val:exmp-1}, the $1$'s in $\Binpy$ represents
information flow from node $0$ to cars $1,2,3$.  We give our first
main result for $Y$ direction motion of a convoy having node $0$
as leader. 
Let 
\begin{equation} \label{equ:y:dir:leader:inp:auto:part}
\hat{y} := \begin{bmatrix}
           y & v_y
          \end{bmatrix}^T \quad \mbox{ and } \quad
\Gamma^y :=  
\begin{bmatrix}\mathrm{0} & \mathrm{I}\\
-k\ \tilde{\Lap}^y & -b\ \tilde{\Lap}^y
\end{bmatrix}
\end{equation}

\begin{theo} \label{theo:main:y:multilane:diff:weight} Consider
a weighted directed graph such that the total weight ($W$) across
all incoming edges is the same for each node. Then the autonomous
system  $\dot{\hat{y}} = \Gamma^y \hat{y}$ achieves an
asymptotically stable equilibrium point at the origin ($0\in\mathbb{R}^{2n}$).
Moreover if the leader velocity $v_{y0}$ is constant, then:  
\begin{enumerate}
\item $v_{yi}\rightarrow v_{y0}\ \mbox{as }t\rightarrow\infty\;\forall\ i\in\{1,\hdots n\}$ 
\item $|y_{i}(t)-y_{j}(t)|\rightarrow0$ as $t\rightarrow\infty$ for all
$i,j$ in the same level. 
\item At equilibrium, the relative spacing between cars in two consecutive
levels is $g_{y}/W$. 
\end{enumerate}
\end{theo}

Before proving the theorems we note that the weighted Laplacian
has a zero row corresponding to the phantom leader, has non-negative
real eigenvalues and $0$ is a simple eigenvalue, with $\begin{bmatrix}1 & 0 & \hdots & 0\end{bmatrix}$
as its left eigenvector, and the Laplacian right nullspace consists
of the vector $[1\ 1\ \hdots\ 1]^{T}$. Additionally the following
result holds.

\begin{lemma}\label{lem:lap:prop:uni} Under the numbering scheme
in Assumption \ref{assump:car:num}, the influence graph has no cycles
and the Laplacian ($\Lapy$) has a lower triangular structure. Moreover,
the diagonal entries of $\Lapy$ will be same for all rows excluding the
rows representing level one and node $0$. 

\label{pro:eig:left} 

\end{lemma}

\prthm{theo:main:y:multilane:diff:weight}  

We first analyze the asymptotic stability of the autonomous part of
the system in \eqref{equ:control:law:y:direction:leader:inp:main}. From
\cite{RenAtk07} we get $Re(\lambda_{i}(\Gamma))<0\ \forall\ i,\ \forall\ b,\; k>0,$
if $ \lambda_{j}(-\tilde{\Lap}^y)<0\ \ \forall\ j$ and $\lambda_{j}(\tilde{\Lap}^y)\in\Real\ \forall\ j$,
where $Re(.)$ denotes the real part. The equilibrium point for the
autonomous part of the system in \eqref{equ:control:law:y:direction:leader:inp:main}
is $v_{y}=0$ and $-\tilde{\Lap}^yy=g_{y}$. 

Next we analyze the system in  \eqref{equ:control:law:y:direction:leader:inp:main}
at equilibrium with fixed velocity $v_{0}$ of node $0$ as input.
 The required spacing can be computed from. 
\begin{equation}
-\Lapy y=g_{y}\begin{bmatrix}0\\
\mathbf{1}
\end{bmatrix}\label{equ:dis:y:we}
\end{equation}
where $\mathbf{1}\in\Real^{n}$ for $n+1$ nodes.
Now we use induction. Cars in the first level
will have one link connecting to the node $0$ by  A2.
For any car $i$ in the first level from \eqref{equ:dis:y:we} we
get, $y_{0}=y_{i}+g_{y}/W$. This will ensure that all cars in the
first level have the same $y$ coordinate.

Consider some car $i$ in level $\mathcal{I}$. Let cars in the level
above be denoted by $j_{1},j_{2},\hdots,j_{n}$. Let $y$ coordinates
of the cars $j_{1},\hdots,j_{n}$ be same. Let the weights on the
links from car $j_{1}$ to car $i$ be denoted by $w_{ij_{1}}$ and
so on. The $i^{th}$ row in \eqref{equ:dis:y:we} can be expressed
as follows: $w_{ij_{1}}y_{j_{1}}+w_{ij_{2}}y_{j_{2}}\hdots+w_{ij_{n}}y_{j_{n}}=g_{y}+Wy_{i}.$
From induction hypothesis, $y_{j}=y_{i}+g_{y}/W$. This completes
the proof. \EP

Clearly, for unweighted influence graphs, the conditions for Theorem \ref{theo:main:y:multilane:diff:weight}
reduces to each car having the same indegree (say $M$), thereby guaranteeing
the relative spacing between two consecutive levels to be  $g_{y}/M$.

\section{Analysis of $X$-dynamics}

\label{subsec:x:axis:mot:uni}\label{sec:bidirec:info:flow}

In this section we analyze the vehicle motions in $X$-direction under
the assumptions stated in Section \ref{sub:sec:assump:for:x}.  The analysis,
while being similar to the $Y$-dynamics, differs in some crucial
features. One of the main differences arise since the influence graph
can be bidirectional (see Figure \ref{fig:bidir:cars_multi_lane_x_3_cars}):
in other words, two drivers can simultaneously look towards each other
and decide on their $X$-control. Secondly, recall that the $X$-velocities
of the pseudo (road boundary) cars (e.g. cars 1, 5, 9) are zero by
assumption (e.g. $v_{x1}=v_{x5}=v_{x9}=0$), and hence their $X$-positions
are always constant ($x_{1}=x_{5}=x_{9}=$ constant). Hence when we
rewrite \eqref{equ:control:law:x:axis} with node $1$ velocity and
position as external input,
\begin{equation}
\begin{bmatrix}\dot{x}\\
\dot{v}_{x}
\end{bmatrix}=\begin{bmatrix}\mathit{0} & \mathrm{I}\\
-k_x\ \tilde{\Lap}^x & -b_x\ \tilde{\Lap}^x
\end{bmatrix}\begin{bmatrix}{x}\\
{v_{x}}
\end{bmatrix}+k_xg_{x}\begin{bmatrix}\mathbf{0}\\
C
\end{bmatrix}+\begin{bmatrix}\mathbf{0}\\
\Binpx
\end{bmatrix}\begin{bmatrix}0\\
k_xx_{1}
\end{bmatrix}\label{equ:control:law:x:axis:leader:inp}
\end{equation}
where $x,v_{x}\in\Real^{n-1}$, $\Binpx$ contains the columns from
$\Lapx$ which denote the links between the leader node 1  and the remaining cars
in the formation. We assume that every car wants to position themselves
at a distance $g_{x}$ from both adjoining cars in the same 'level'.
Let,
\begin{equation} \label{equ:x:dir:leader:inp:auto:part}
\hat{x} := \begin{bmatrix}
           x & v_x
          \end{bmatrix}^T \quad \mbox{and} \quad          
\Gamma^x :=  
\begin{bmatrix}\mathrm{0} & \mathrm{I}\\
-k_x\ \tilde{\Lap}^x & -b_x\ \tilde{\Lap}^x
\end{bmatrix}
\end{equation}
\begin{theo} \label{theo:main:res:bidir:x:axis} Under the above
assumptions, there exists $b$, $k$ such that the autonomous system $\dot{\hat{x}}=\Gamma^x \hat{x}$
  asymptotically achieves
equilibrium at the origin ($0\in\mathbb{R}^{2n})$.
Moreover, 
\begin{enumerate}
\item $v_{xi}(t)\rightarrow0\ \mbox{as }t\rightarrow\infty\;\forall\ i\in\{1,\hdots,n\}$. 
\item There exist $C$ such that $|x_{i}-x_{i+1}|\rightarrow g_{x}$ as $t\rightarrow\infty$
for $i,i+1$ in same level. 
\item The $C$ achieving (2) can be computed locally.
\end{enumerate}
\end{theo}

For obtaining a spacing of $g_{x}$ between cars in the same level,
we impose the additional constraints for cars $i$ and $i+1$ in the
same level (denoted by $\mathcal{I}$) as follows: 
\begin{align}
x_{i+1}-x_{i}=g_{x}\quad\forall\ i,\ i+1\in\mathcal{I}\label{equ:extra:req:bidirec}
\end{align}

\prthm{theo:main:res:bidir:x:axis} The stability condition on constants
$b$ and $k$ in control law \eqref{equ:control:law:x:axis:leader:inp}
can be easily obtained from \cite{RenAtk07}. Claim (1) follows immediately.

Next we show the existence of $C=[0\ z_{1}\ \hdots\ z_{n}]^{T}$ in
\eqref{equ:control:law:x:axis:leader:inp} for cars in level $\mathcal{I}$.
The same arguments can be repeated for other levels. Let $m$ and
$\tilde{m}$ be the total number of cars in level $\mathcal{I}$ and
in the level above $(\mathcal{I}-1)$. Let the cars in level $\mathcal{I}$
be linked to cars $j_{k}$ with weights given by $w_{ij_{k}}\ \forall\ k\in(\mathcal{I}-1)$.
Then the following equation can be obtained from \eqref{equ:control:law:x:axis}
at equilibrium: 
\begin{equation}
\Lapx x=-g_{x}C\label{equ:part:x:bidirec}
\end{equation}
Choosing rows corresponding to cars in level $\mathcal{I}$ from \eqref{equ:part:x:bidirec},
and combining with \eqref{equ:extra:req:bidirec}, we get

\begin{align}
\left[\scalemath{0.64}{\begin{array}{ccccccccc}
1 & 0 & \hdots & -1 & 0 & \hdots & 0 & \hdots & 0\\
0 & w_{ij_{1}} & \hdots & w{ii} & w_{ii+1} & -1 & 0 & \hdots & 0\\
0 & w_{i+1j_{1}} & \hdots & w{i+1i+1} & w_{i+1i+2} & 0 & -1 & \hdots & 0\\
\vdots & \vdots & \vdots & \vdots & \vdots & \vdots & \vdots & \vdots & \vdots\\
\hdashline0 & \hdots & -1 & 1 & \hdots & 0 & 0 & 0 & 0\\
0 & \hdots & 0 & -1 & 1 & 0 & 0 & 0 & 0\\
\vdots & \vdots & \vdots & \vdots & \vdots & \vdots & \vdots & \vdots & \vdots
\end{array}}\right]\left[\begin{smallmatrix}x_{1}\\
\vdots\\
x_{m+\tilde{m}}\\
g_{x}z_{1}\\
\vdots\\
g_{x}z_{m+\tilde{m}}
\end{smallmatrix}\right]=\left[\begin{smallmatrix}0\\
\vdots\\
0\\
g_{x}\\
\vdots\\
g_{x}
\end{smallmatrix}\right]\label{equ:x:part:lap:bidir}
\end{align}
We show the existence and uniqueness of constants $z_{i}$ using rank
conditions. Rearranging columns of the augmented matrix from \eqref{equ:x:part:lap:bidir}
we get: 
\begin{equation}
\left[\begin{smallmatrix}1 & 0 & 0 & \hdots & 0 & -1 & 0 & \hdots & 0 & 0\\
0 & -1 & 0 & \hdots & w_{ij_{1}} & -w_{ii} & w_{ii+1} & 0 & \hdots & 0\\
0 & 0 & -1 & 0 & \hdots & w_{i+1j_{1}} & \hdots & w_{i+1i+1} & \hdots & 0\\
\vdots & \vdots & \vdots & \vdots & \vdots & \vdots & \vdots & \vdots & \vdots & \vdots\\
0 & 0 & \hdots & 0 & -1 & 1 & \hdots & 0 & 0 & g_{x}\\
\vdots & \vdots & \vdots & \vdots & \vdots & \vdots & \vdots & \vdots & \vdots & \vdots\\
0 & 0 & \hdots & 0 & 0 & -1 & 1 & 0 & \hdots & g_{x}
\end{smallmatrix}\right]\label{equ:x:part:lap:bidir:rhs}
\end{equation}

The matrix in \eqref{equ:x:part:lap:bidir:rhs} is in row echelon
form with unity pivot elements. Thus \eqref{equ:x:part:lap:bidir:rhs}
is full row rank matrix. This guarantees the existence of at least
one solution ($z_{i}$) giving the desired equilibrium spacing. 

\label{rem:loc:comp} The local/distributed nature of the $C$ computation
can be verified by considering \eqref{equ:x:part:lap:bidir} for car
$i$: $w_{ij_{1}}x_{j1}+\hdots+w_{ij_{n}}x_{jn}-w_{ii}x_{i}=g_{x}z_{i}$
and $x_{i+1}-x_{i}=g_{x}$. Clearly these equations only require information
from the immediate $x$-neighbours of the car $i$.  \EP

\begin{remark}\label{rem:xf:def}
At the expense of some new notation, an explicit formula for $C$
can be given easily. \label{prop:xf:deter:x:axis} Let $x_{f}\in\Real^{n}$
be such that $x_{f}g_{x}\in\mathbb{R}^{n}$ give the $x$-coordinates
of all the cars in the formation. For example in Figure \ref{fig:bidir:cars_multi_lane_x_3_cars}
if we want all cars in the same level to have maintain a spacing of
$g_{x}$ from each other the corresponding $x_{f}$ should be, $x_{f}=[0\ 1\ 2\ 3\ \hdots0\ 1\ 2\ 3]^{T}$.
Then it is easy to verify that the constant vector $C$ in control
law \eqref{equ:control:law:x:axis} is given by $C=\Lap x_{f}$.  
\end{remark}

\begin{remark}
Though the above results have assumed that influences percolate across
atmost one level, this assumption can be easily extended to multi-level
influence graphs. Such a requirement might arise, e.g. for a driver
looking in the rear view mirror while taking turns to avoid cars behind
him. 
\end{remark}

\section{Time varying graphs}

\label{sec:time:vary:graph} In this section we analyze situations
where the influence graph changes over time, due to relative motion
of the vehicles during various traffic events, e.g., road widening
or traffic signals. During road widening scenario, cars moving in
a convoy may spread out. As the cars approach traffic signals the
cars in the formation may come closer. As this motion takes place,
cars may move in and out of influence cones of other agents, thereby
changing the influence graphs, the transition matrix and the Laplacian.
Hence  \eqref{equ:control:law:y:direction} and \eqref{equ:control:law:x:axis} become switched systems. We assume,
quite reasonably, that this switchings occur after finite intervals
of time. Let $\mathit{P}$ be a finite index set: $\mathit{P}=\{1,2,\hdots m\}$.
Let $\sigma$ be piece wise constant switching signal $\sigma:[0,\infty)\to\mathit{P}$,
$\Gammasig$ be the transition matrix and $\Lapsig$ the graph of
the Laplacian of the system in \eqref{equ:control:law:y:direction}
and \eqref{equ:control:law:x:axis} corresponding to $\sigma(t)$.
We analyze the $Y$-dynamics only for lack of space. Similar arguments
work for the $X$- dynamics as well. The switched version of \eqref{equ:control:law:y:direction}
is given by,

\begin{equation}
\begin{bmatrix}\dot{y}\\
\dot{v}_{y}
\end{bmatrix}=\begin{bmatrix}\mathit{0} & \mathrm{I}\\
-k\ \Lapsigt^y & -b\ \Lapsigt^y
\end{bmatrix}\begin{bmatrix}{y}\\
{v_{y}}
\end{bmatrix}-kg_{y}\begin{bmatrix}\mathbf{0}\\
\mathbf{1}
\end{bmatrix}+\begin{bmatrix}\mathbf{0}\\
\Binpy_{\sigma}
\end{bmatrix}\begin{bmatrix}bv_{0}\\
ky_{0}
\end{bmatrix}\label{equ:control:law:y:axis:leader:inp}
\end{equation}
where $\Binpy_{\sigma}\in\Real^{n\times2}$, $y,v_{y}\in\Real^{n}$. Clearly,
$\Lapsigt^y$, $\Lapsig^y$, and $\Binpy_{\sigma}$ changes with the switching
influence graph. In this section we assume that the cars still want
to preserve the same inter-vehicle spacing even though the graph is
changing. Hence the spacing constant $g_{y}$ remains same for all
$\sigma(t)$. We show that this is possible by local re-computation
of the weights assigned to the incoming edges by each car. We now
give the main result. 

\begin{theo}\label{theo:time:vary:y:graph} Under the assumptions
stated in Section \ref{sub:Assumptions-for--y}, the autonomous part of system
\eqref{equ:control:law:y:axis:leader:inp} is globally uniformly asymptotically
stable and \eqref{equ:control:law:y:axis:leader:inp} is BIBO stable.
Moreover, if $v_{y0}$ is constant and the net weight of all incoming
edges for each node be kept same ($W$) for all $\sigma(t)$, then:
\begin{enumerate}
\item $v_{yi}\rightarrow v_{y0}\ \mbox{as }t\rightarrow\infty\;\forall\ i\in\{1,\hdots n\}$ 
\item $|y_{i}(t)-y_{j}(t)|\rightarrow0$ as $t\rightarrow\infty$ for all
$i,j$ in the same level. 
\item At equilibrium, the relative spacing between cars in two consecutive
levels is $g_{y}/W$. 
\end{enumerate}
\end{theo}

\prthm{theo:time:vary:y:graph}{[}Sketch of Proof{]} Stability for
each subsystem in \eqref{equ:control:law:y:axis:leader:inp} was shown
in Theorem \ref{theo:main:y:multilane:diff:weight}. The transition matrix for the system
subject to switching signal $\sigma(t)$ in \eqref{equ:control:law:y:axis:leader:inp}
is given by 
\[
\Gammasigt^y=\begin{bmatrix}0 & I\\
-k\Lapsigt^y & -b\Lapsigt^y
\end{bmatrix}\mbox{ and let }P=\begin{bmatrix}\tilde{I} & 0\\
0 & R
\end{bmatrix},
\]
where $\ R=R^{T}>0,\ \tilde{I}=I_{n\times n}/q,\ q$ is a constant
satisfying ${1}/{kq} < \lambda_{\rm min} (\Lapsig^y) $.
Noting that $\Lapsigt^{yT} + \Lapsigt^y >0$ for $q$ as stated here, it is easy to verify
under the assumptions of Section \ref{sub:Assumptions-for--y}, that $\Gammasigt^{yT}P+P\Gammasigt^y<0$.
It follows \cite{bookdliber} that the autonomous part of \eqref{equ:control:law:y:axis:leader:inp}
is globally uniform asymptotically stable (GUAS). Further it is well known
\cite{MiGe02} that for a linear switched system, GUAS
ensures BIBO stability. Claims (1), (2) and (3) follow in a  similar fashion as in 
Theorem \ref{theo:main:y:multilane:diff:weight}.

\begin{remark} We have assumed that the net weight for each node
remains same. At every switching instant, we assume that each car
can locally redistribute the available weight on its links. With this
the equilibrium point of the autonomous system in \eqref{equ:control:law:y:axis:leader:inp}
remains invariant across switching. \end{remark}

\subsection{Obstacles}

\begin{figure}
\begin{subfigure}[b]{0.16\textwidth} \includegraphics[scale=0.5]{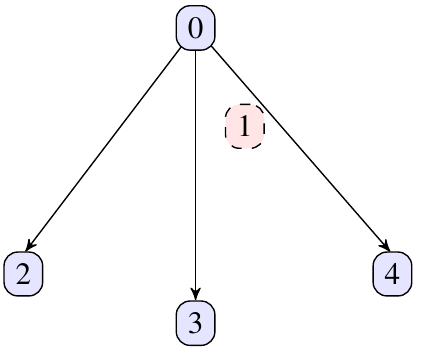}
\caption{Graph before \protect \\
 obstacle is sensed}

\label{fig:obs:basic:1} \end{subfigure} \begin{subfigure}[b]{0.16\textwidth}
\includegraphics[scale=0.5]{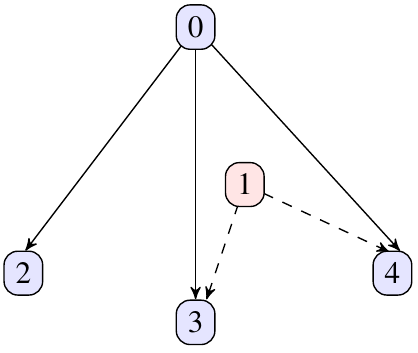} \caption{Graph when \protect \\
 obstacle is sensed}

\label{fig:obs:basic:2} \end{subfigure}\begin{subfigure}[b]{0.16\textwidth}
\includegraphics[scale=0.5]{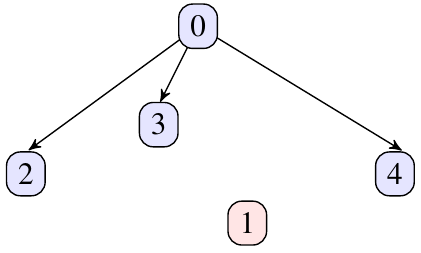} \caption{Graph afer \protect \\
 obstacle has passed}

\label{fig:obs:basic:3} \end{subfigure} \caption{Changes in information flow graph for $Y$ direction when $4$ cars
encounter an obstacle denoted by node $1$}

\label{fig:obs:basic} 
\end{figure}

\label{sec:obs:avoid} Using the switching theory developed above,
stability of inter vehicle spacing in the presence of stationary
obstacles can be ensured. The obstacle is modeled as a stationary
car having zero velocity in both directions and is numbered consecutively
along with other nodes. Figure \ref{fig:obs:basic} shows the transitions
in the influence graph for a typical obstacle crossing. We briefly describe
the dynamics in $Y$-direction. Similar arguments hold for the $X$-dynamics.

Let $v_{0},v_{i}$ and $y_{0},y_{i}$ be the velocity and position
inputs of leader node $0$ and obstacle denoted by node $i$ respectively.
By hypothesis, $v_{i}=0$ and $y_{i}=$ constant. The control law
for motion in $Y$ direction:

\begin{equation}
\begin{bmatrix}\dot{y}\\
\dot{v}_{y}
\end{bmatrix}=\begin{bmatrix}\mathit{0} & \mathrm{I}\\
-k\Lapsigt^y & -b\Lapsigt^y
\end{bmatrix}\begin{bmatrix}{y}\\
{v_{y}}
\end{bmatrix}-kg_{y}\begin{bmatrix}\mathbf{0}\\
\mathbf{1}
\end{bmatrix}+\begin{bmatrix}\mathbf{0}\\
\Binpy_{\sigma}
\end{bmatrix}\begin{bmatrix}bv_{0}\\
ky_{0}\\
bv_{i}\\
ky_{i}
\end{bmatrix}\label{equ:control:y:lead:obs}
\end{equation}
where $B_{\sigma}\in\Real^{n-1\times4}$, $y,v_{y}\in\Real^{n-1}$.
As usual, $B_{\sigma}$ identifies the (switching) links between leader
node $0$ and the remaining cars (except node $i$) and between node
$i$ and the remaining cars in the formation. 

\begin{corollary}
Suppose all cars influenced by the obstacle
are also influenced simultaneously by at least one more car in the
formation. Then Theorem \ref{theo:time:vary:y:graph} holds for \eqref{equ:control:y:lead:obs}.   
\end{corollary}

\subsection{Lane Change}

\label{sec:lane:change}

\begin{figure}
\centering 
\begin{minipage}{0.25\textwidth} 
\centering \includegraphics[scale=0.7]{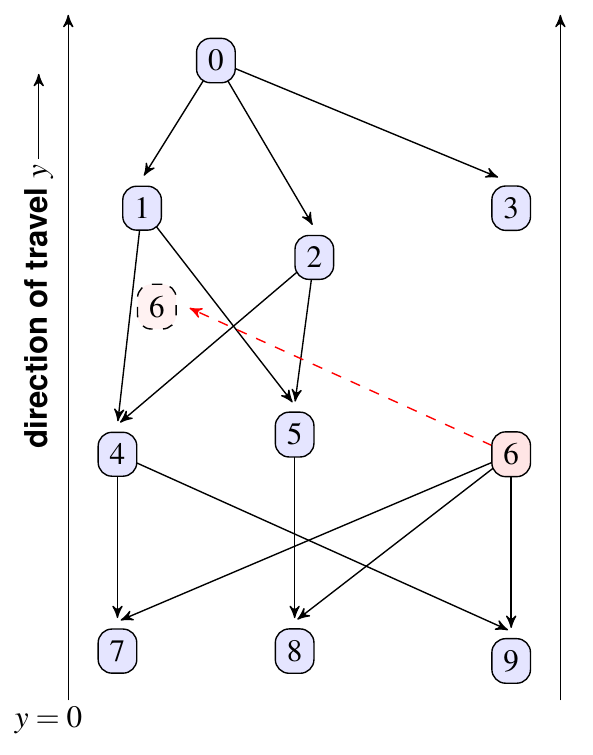}
\caption{{Information flow in $Y$ axis with car $6$ changing\\  lanes with 
trajectory shown\\ by the  red dashed line } } \label{fig:lane:change:graph:multi:y}
\end{minipage}%
\begin{minipage}{0.25\textwidth}
\centering
 \includegraphics[scale=0.6]{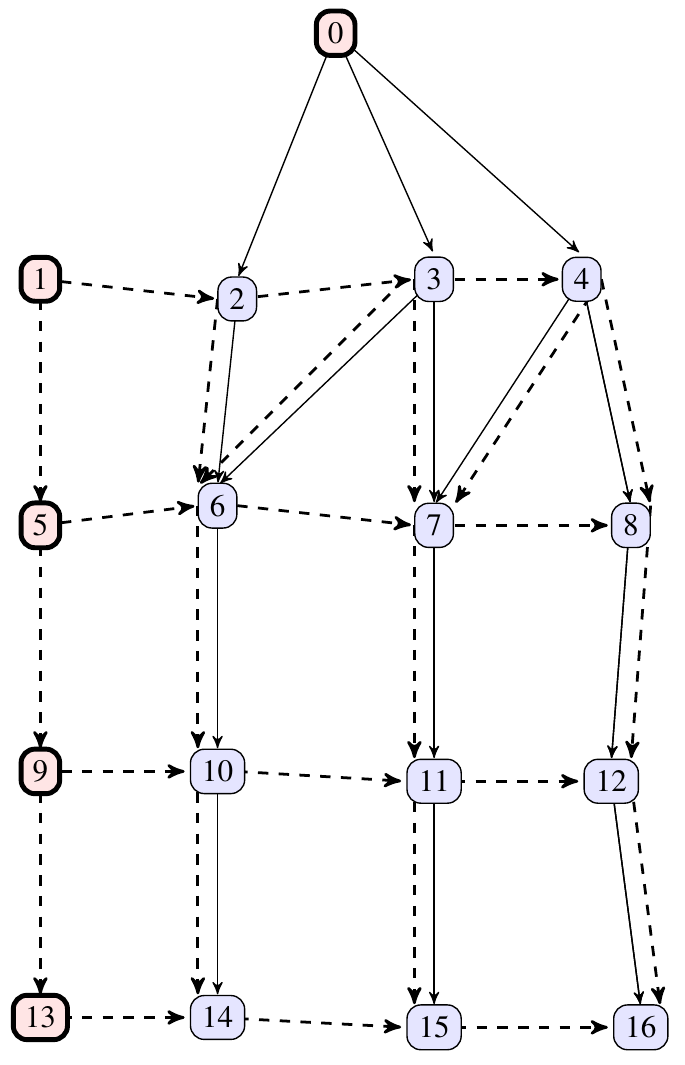}
 \caption{Influence graphs based on  initial conditions in simulations   
 Dashed lines indicate $X$ graph.  Solid lines represent $Y$ graph}\label{fig:infl:graph:sim:ini}
\end{minipage}
\vspace{-15pt}
\end{figure}

In lane changing scenarios we assume that a particular agent decides
to change his $X$ position arbitrarily within the formation. We assume that the maneuver
takes finite time to complete. This is illustrated in Figure \ref{fig:lane:change:graph:multi:y}
where car $6$ is changing lanes within the formation. Suppose car
$i$ is changing lanes. We assume car $i$ to be an external input.
The $Y$-axis dynamics is given by: 
\begin{equation}
\begin{bmatrix}\dot{\tilde{y}}\\
\dot{v_{y}}
\end{bmatrix}=\begin{bmatrix}\mathrm{0} & \mathrm{I}\\
-k\ \Lapsigt^y & -b\ \Lapsigt^y
\end{bmatrix}\begin{bmatrix}{\tilde{y}}\\
{v_{y}}
\end{bmatrix}+\begin{bmatrix}\mathbf{0}\\
\Binpy
\end{bmatrix}\begin{bmatrix}v_{0}\\
\tilde{y}_{0}\\
v_{i}\\
\tilde{y}_{i}
\end{bmatrix}\label{equ:control:law:y:lane:change}
\end{equation}
where $\tilde{y},\dot{\tilde{y}},v_{y},\dot{v_{y}}\in\mathbb{R}^{n-1},\ \tilde{y}=y+g_{y}$.
As long as a directed spanning tree continue to exist during the lane
change, clearly Theorem \ref{theo:time:vary:y:graph} continues to
hold for \eqref{equ:control:law:y:lane:change}. Similar analysis holds for the $X$  dynamics but is 
not included for lack of space.

\section{Stability under impulse effects}

\label{sec:stab:impulse:effect}

In previous sections the individual cars were kept equidistant from
each other using the constants $g_{y}$, $g_{x}$, $x_{f}$ (and $C$).
However the cars might want to change the equilibrium spacing in a
variety of situations, such as, when the road narrows/widens and we
want cars to come closer/further or some rogue drivers want to change
their position in the formation arbitrarily. Individual cars can decide
on a new $g_{y}$, $x_{f}$ or $C$ suddenly, introducing discontinuities
in \eqref{equ:control:law:y:direction} and \eqref{equ:control:law:x:axis}. Such discontinuities are known
as \emph{impulse effects} \cite{bookLaBaSi}. For simplicity
we assume that the influence graph remains invariant and the velocity
vectors are continuous for $t\in[0,\infty)$. However, time varying
graphs can also be handled easily.

The autonomous part of the system in \eqref{equ:control:law:y:direction:leader:inp:main} with
impulse effects at fixed time instants $\tau_{k}$ (assume $0=\tau_{0}<\tau_{1}<\tau_{2}<\hdots<\lim_{k\to\infty}\tau_{k}=\infty$)
can be characterised as follows,
\begin{equation}
\begin{bmatrix}\dot{\tilde{y}}\\
\dot{v_{y}}
\end{bmatrix}=\begin{bmatrix}\mathrm{0} & \mathrm{I}\\
-k\ \tilde{\Lap} & -b\ \tilde{\Lap}
\end{bmatrix}\begin{bmatrix}{\tilde{y}}\\
{v_{y}}
\end{bmatrix}\quad t\ne\tau_{k}\label{equ:ctrl:law:y:imp:effec}
\end{equation}
where $\tilde{y}=y+g_{y}$. Let the cars decide to change instanteneously
the spacing constant from $g_{y}(t)$ to $g_{y}(t^{+})=g_{y}(t)+\tilde{g_{y}}$
at $t=\tau_{k}$. This creates a discontinuous change in the position
state vector $\tilde{y}(t^{+})=\tilde{y}(t)+\Delta\tilde{y}$ where
$\Delta\tilde{y}={\normalcolor \textcolor{red}{{\normalcolor \tilde{g}_{y}}}}\quad \mbox{at } t=\tau_{k}$
Under these assumptions, the existence
and uniqueness of solutions for the system in \eqref{equ:ctrl:law:y:imp:effec}
is ensured \cite[Theorem 1.6.2]{bookYang}. 

\begin{theo}\label{theo:impulse:pheno:stab} Suppose $\tilde{\Lap}$
remains unchanged for $t\in[0,\infty)$ and $v_{y}(t^{+})=v_{y}(t)$.
Then \eqref{equ:ctrl:law:y:imp:effec} is exponentially stable if
$g_{y}(t^{+})$ is chosen such that $||\tilde{y}(t^{+})||\le||\tilde{y}(t)||$. 

\end{theo}

\prthm{theo:impulse:pheno:stab} Let $\hat{y}:=\begin{bmatrix}\tilde{y} & v_{y}\end{bmatrix}^{T}$.
The common Lyapunov function at time $t\ne\tau_{k}$ is given by,
$V(t,\hat{y})=\hat{y}^{T}\begin{bmatrix}I/q & \mathbf{0}\\
\mathbf{0} & I
\end{bmatrix}\hat{y}$ where $\frac{1}{qk}<\lambda_{{\rm min}}(\Lap)$. At the switching
instant $t=\tau_{k}$, $V(t^{+},\hat{y}+\hat{g}_{y})=(\hat{y}+\hat{g}_{y})^{T}\begin{bmatrix}\hat{I} & \mathbf{0}\\
\mathbf{0} & I
\end{bmatrix}(\hat{y}+\hat{g}_{y})$ where $\hat{g}_{y}:=\begin{bmatrix}\tilde{g_{y}} & \mathbf{0}\end{bmatrix}^{T},\ \hat{g}_{y}\in\Real^{2n}.$
The $\mathbf{0}\in\hat{g}_{y}$ exists because we have assumed that
the velocity vector is a continuous for $t\in[0,\infty)$.

From \cite{bookLaBaSi}, and \cite[Theorem 4.14]{bookYang}: 
$
V(t,\hat{y})\ge V(t^{+},\hat{y}+\hat{g}_{y})$
guarantees exponential stability. It is easy to be verified that this condition is satisfied if
 $||\tilde{y}(t)||\ge||\tilde{y}(t^{+})||$
This completes the proof. \EP

A similar result holds for impulse effects in $X$-dynamics.

%
%
%
%
%

\section{Simulations}\label{sec:simul}%
In this section we present some numerical simulations. In our numerical simulations 
we use the scenarios of  changing formations,
obstacle avoidance, and lane change. We use the initial conditions
depicted by graphs in Figure \ref{fig:infl:graph:sim:ini}
for $Y$ and $X$ axis motion respectively. Unidirectional communication
for $X$ and $Y$ axis is assumed with an influence level of one.
Cars $1,5,9,13$ indicate the boundary of the road in Figure \ref{fig:infl:graph:sim:ini}.
The maximum number of cars in each level is $4$. The angle of view
for $Y$ axis is a $120$\textdegree{}cone and for $X$ axis it is a $180$\textdegree{}angle.
The system parameters in \eqref{equ:control:law:y:direction} and
\eqref{equ:control:law:x:axis} are set as follows: $b=0.4$ and $k=0.001$.
The vertical spacing constant between consecutive levels is $g_{y}=50$
and horizontal spacing for cars in the same level is $g_{x}=30$.



 \begin{figure}
 \centering
 \includegraphics[scale=0.4]{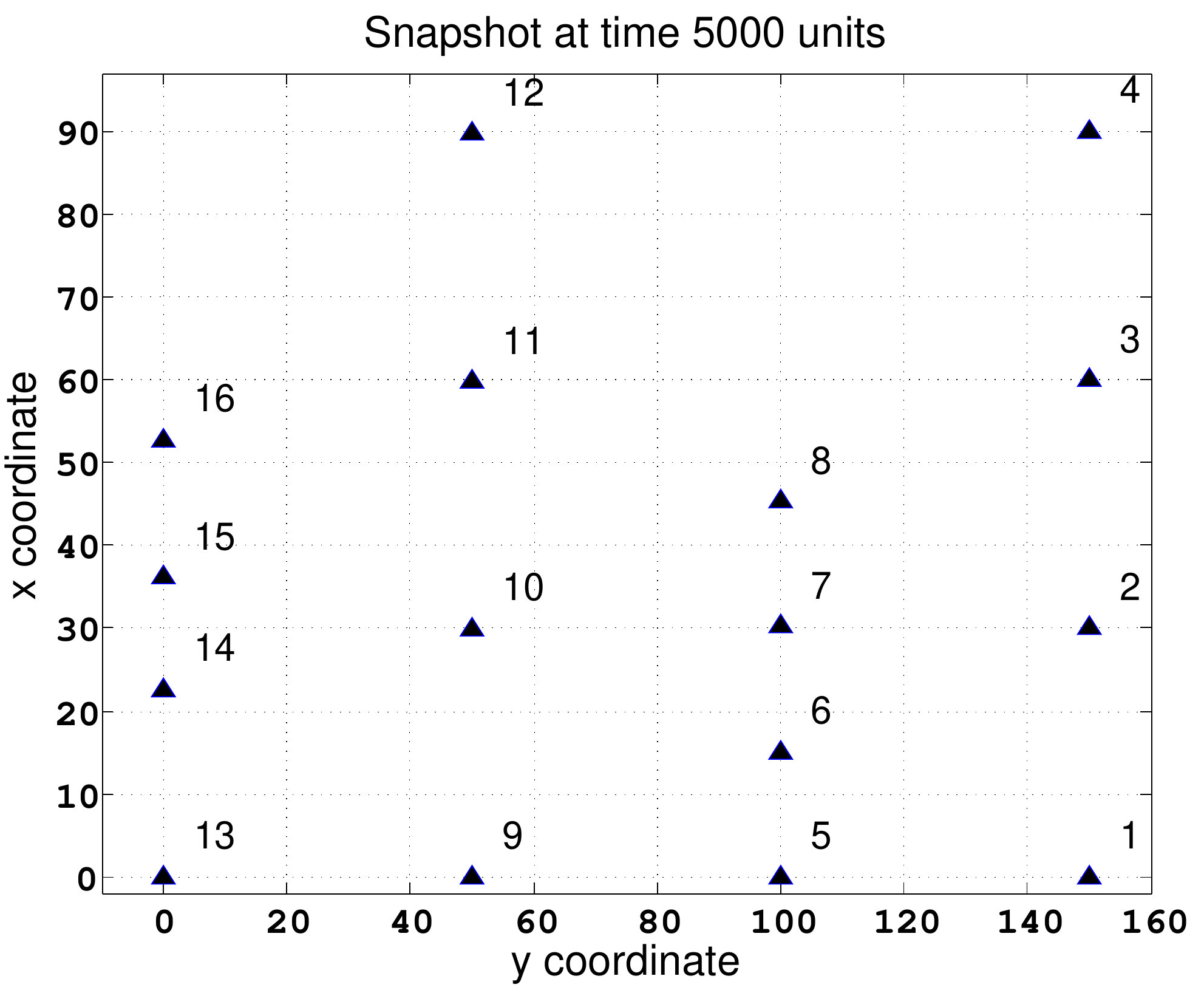}
\caption{The $X$ vs $Y$ coordinates of cars are indicated at a time instant just before changing formations. The system is 
at equilibrium at this time}
\label{fig:chng:form:1}
\vspace{-10pt}
\end{figure}

\begin{figure} 
 \centering
 \includegraphics[scale=0.4]{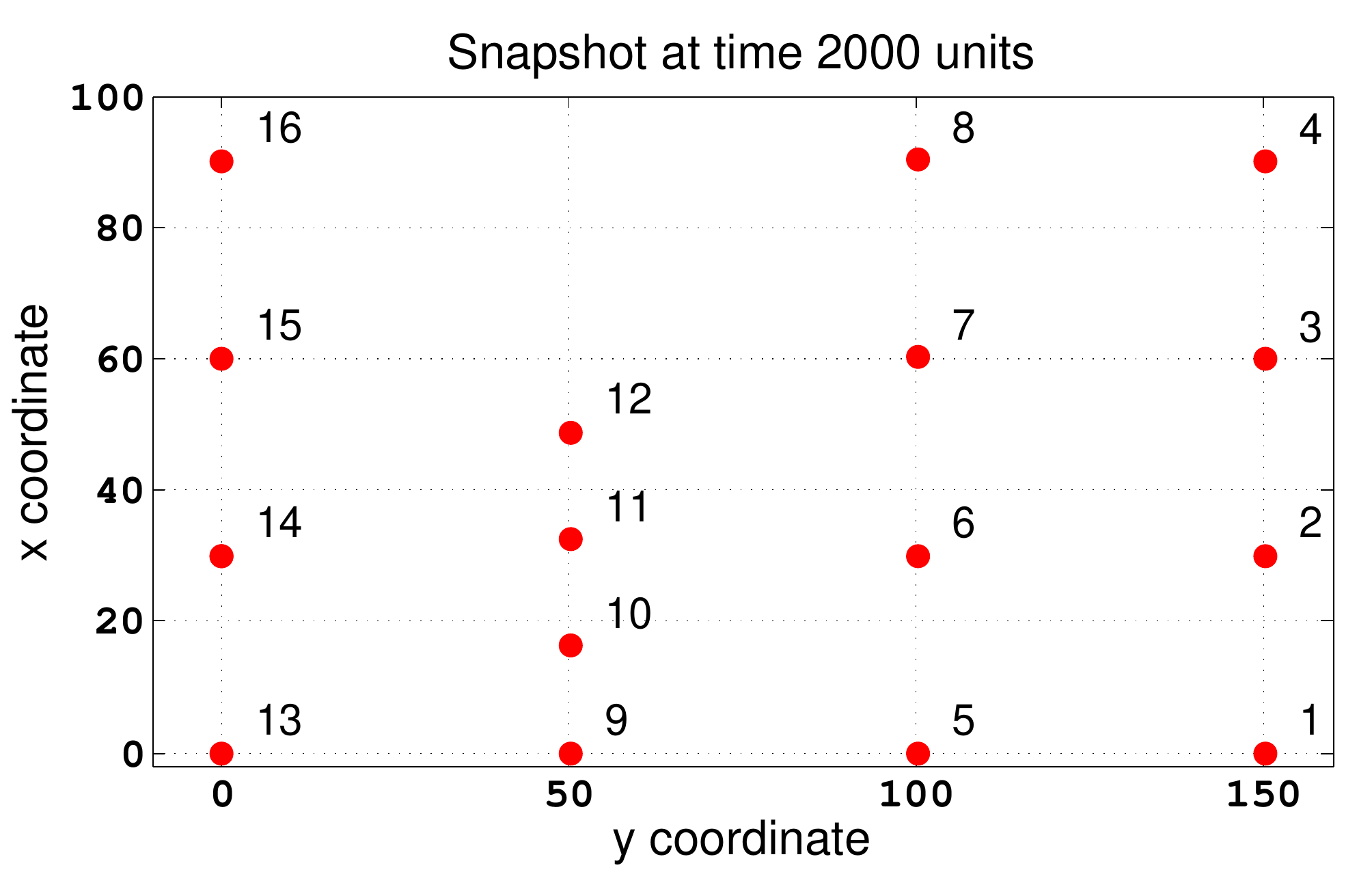}
 \caption{The $X$ vs $Y$ coordinates of cars are indicated at a time instant after formation has changed and the 
 system has settled at equilibrium}
\label{fig:chng:form:2}
\vspace{-10pt}
\end{figure}

%


\begin{figure}
\centering
\includegraphics[scale=0.3]{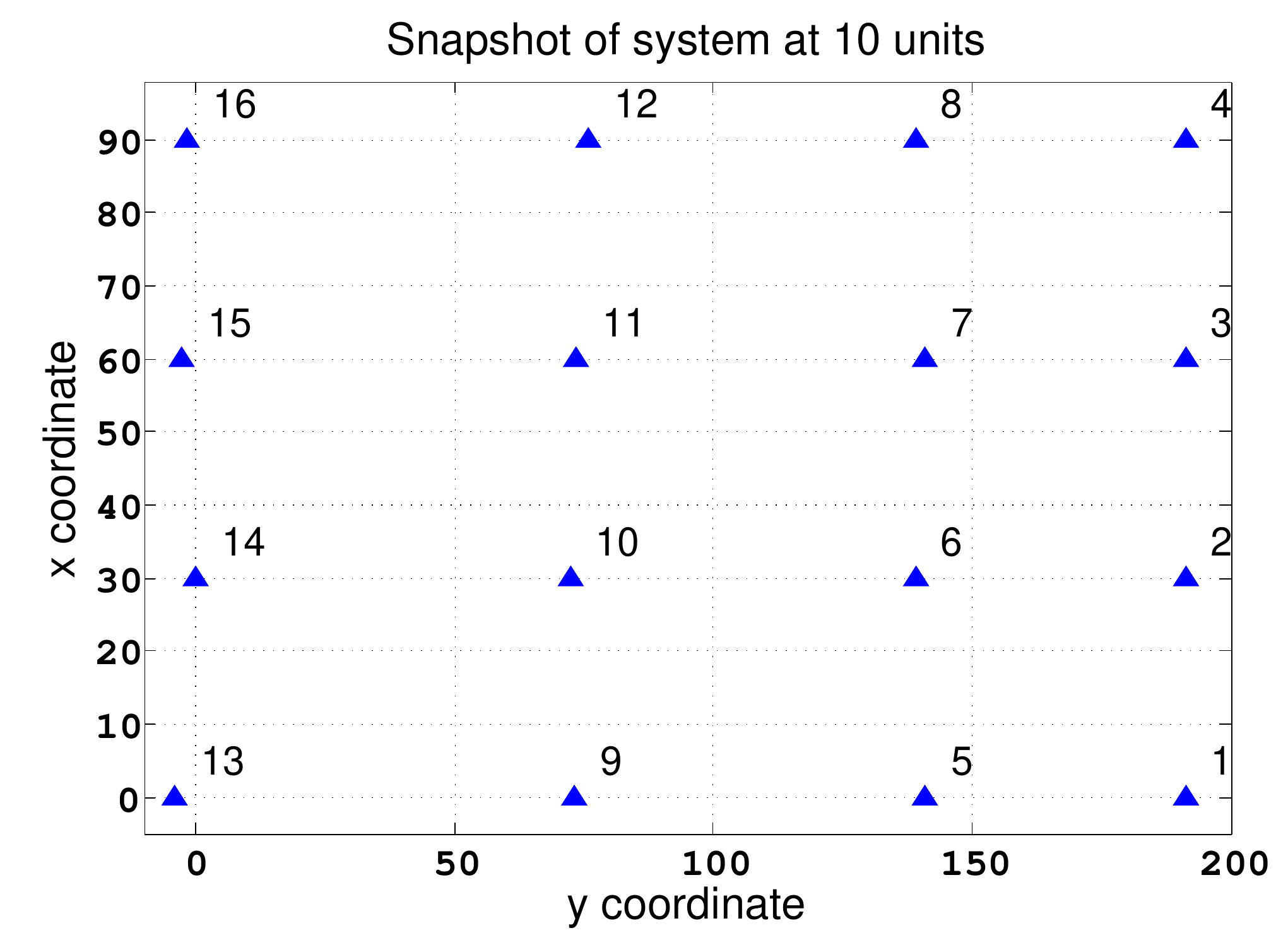}
\caption{Figure shows the formation at the instant before car $6$ changes lanes}
\label{fig:lane:change:1}
\vspace{-15pt}
\end{figure}%
\begin{figure}
\centering
\includegraphics[scale=0.3]{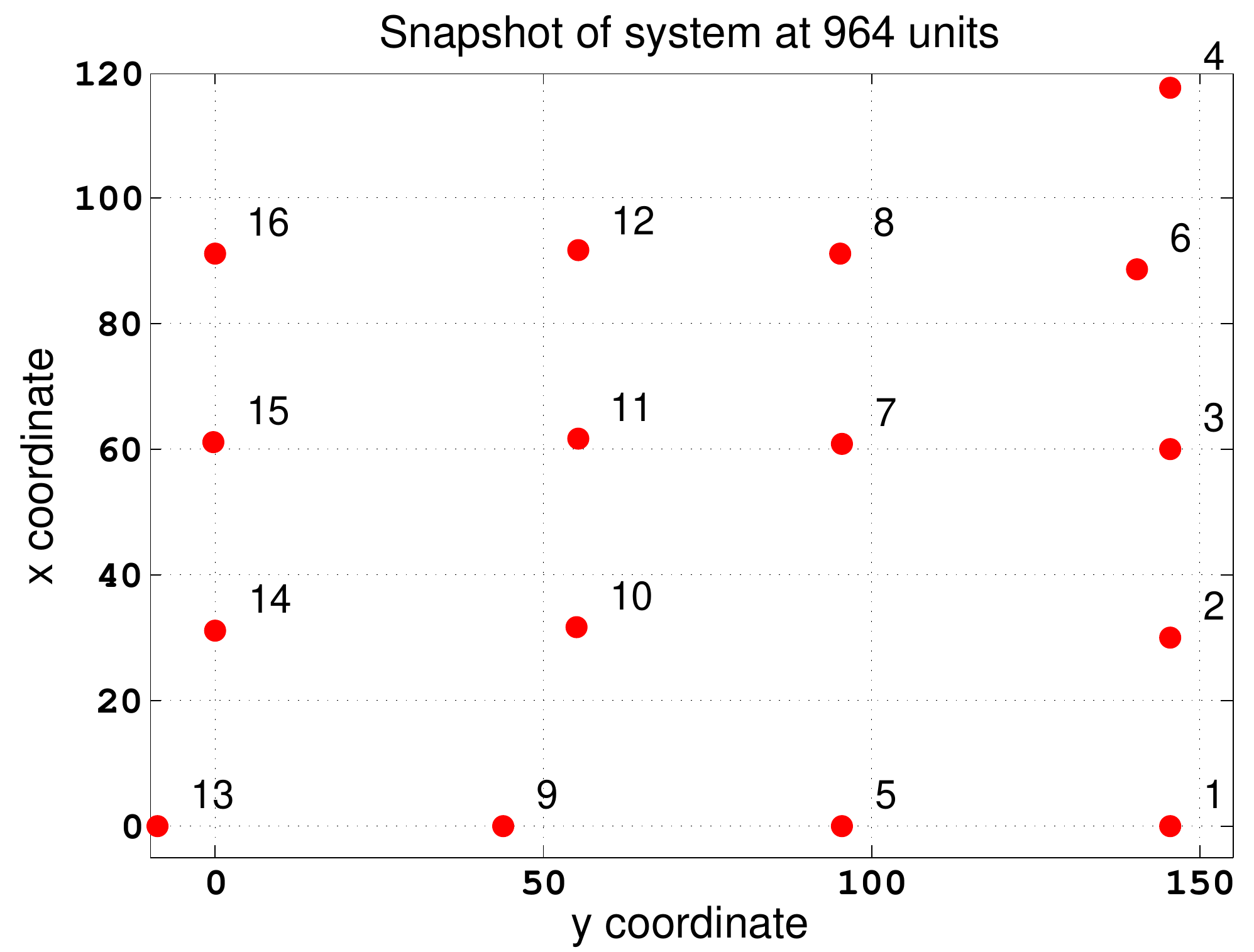}
\caption{Figure shows the formation at the instant after car $6$ has changed lanes}
\label{fig:lane:change:2}
\vspace{-15pt}
\end{figure}%
\subsection{Changing formations}%
When the equilibrium point of $X$ or $Y$ dynamics is changed we observe a change in the formation.
We change vector $x_{f}$ in Remark \ref{rem:xf:def}
at different time instants to attain differing spacing between cars in the same level.
Changing $x_{f}$ introduces impulse effects in the system in \eqref{equ:control:law:y:direction}.
Figures \ref{fig:chng:form:1} and \ref{fig:chng:form:2} demonstrates this effect. 
Cars $1$ to  $4$ act as references and are spaced $30$ units from each other.
 Nodes $1,5,9,13$ represent the boundary. 
  The  `\textcolor{red}{$\bullet$}' and  `\textcolor{blue}{$\blacktriangle$}'
  represents
  the system a snapshot of the system at time $2000$ 
  and $5000$ units respectively. The corresponding values of $x_f$ for which the figures are obtained is as follows.
  For time units of $5000$ units, $x_{f}^{T}=[\ 0\ 1\ 2\ 3\ 0\ 0.5\ 1\ 1.5\ 0\ 1\ 2\ 3\ 0\ 0.5\ 1\ 1.5 ]$. 
For time units of $2000$ units, 
 $x_{f}^{T}=[\ 0\ 1\ 2\ 3\ 0\ 1\ 2\ 3\  0\ 0.5\ 1\ 1.5\  0\ 1\ 2\ 3\ ]$.



%
%
%
%
%
\begin{figure}
\centering
 \includegraphics[scale=0.35]{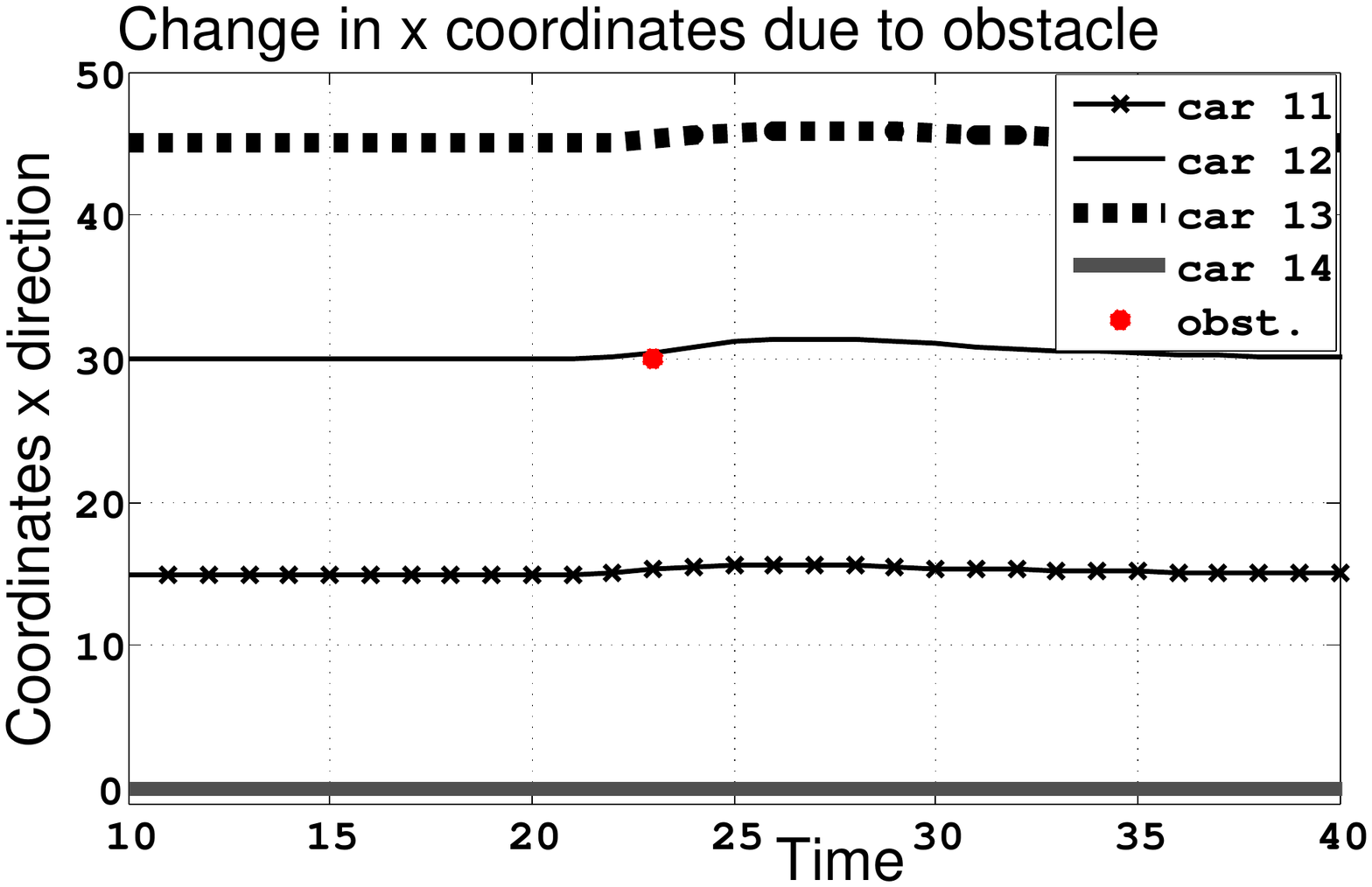}
\caption{Trajectories of cars in $X$ direction changing due to the obstacle}
\label{fig:obs:x:cord:first} 
\vspace{-10pt}
\end{figure}
%
%
%
 \begin{figure}
 \centering
\includegraphics[scale=0.4]{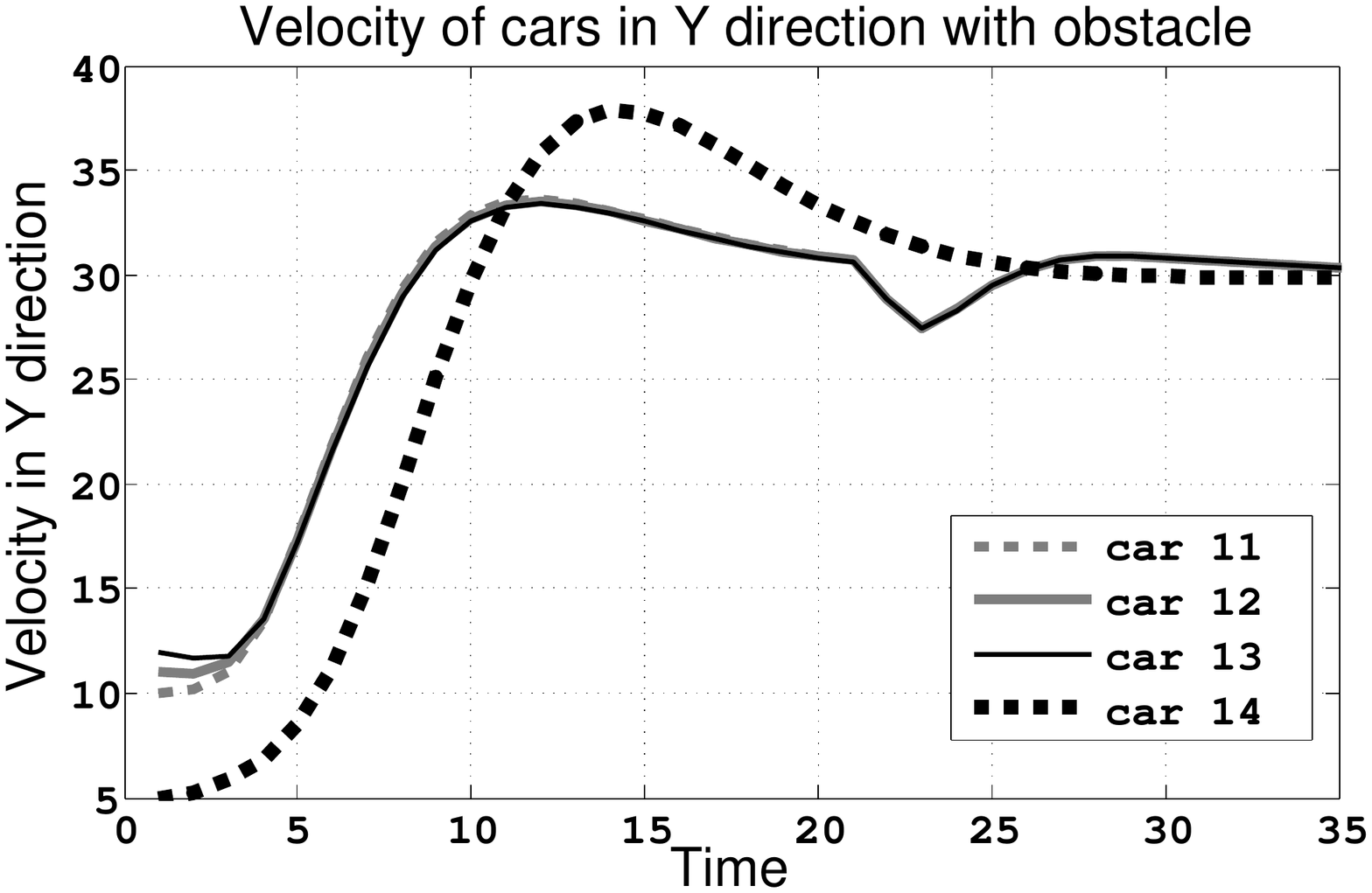}
\caption{Velocities in $Y$ direction are plotted as function of time. The
deviation is due to the obstacle}
\label{fig:obs:y:velo:first} 
\vspace{-15pt}
\end{figure}
\subsection{Obstacle avoidance}
Here we show the impact of a stationary obstacle on the convoy. In
our simulation the obstacle  affects all cars below cars in level $2$, i.e.
cars $10$ onwards. Figure \ref{fig:obs:x:cord:first} plots deviation
of $x$ coordinates with respect to time. In Figure \ref{fig:obs:x:cord:first}
the red star denotes the stationary obstacle. Car $12$ goes around
the obstacle. This causes a change in position and velocity of cars
$11$ and $13$ in the same level. Note that car $11$ has moved
towards the obstacle, this is caused due to the control law forcing
car $11$ to maintain a fixed distance with respect car $12$ and
the obstacle.
 
  Figure \ref{fig:obs:y:velo:first} shows the change in $Y$ velocities
  due to the obstacle.The black dotted trajectory is the reference 
  trajectory in the absence of an obstacle. As cars $11$ to $13$ sense the 
  obstacle their velocities drop as shown in the Figure. This drop causes a 
  deviation in the $y$ coordinates of the cars.
We now present the case for a particular car arbitrarily changing $x$ coordinates within
the formation. 

%
%
%
%
%
%
%
%
%
\subsection{Lane change}
For this example, car $6$ changes lanes from the  left to
the right in the formation. In doing so
car $6$ has moved from one level below to one level above in the
formation. Figure \ref{fig:lane:change:1} shows the formation before lane change.
As car $6$ changes lane other cars are also affected. This can be seen in 
Figure \ref{fig:lane:change:2}	in which cars $7,8$ are moving closer  to car $5$.
Cars $10,11$ are moving closer to cars  $5,7,8$ in $Y$ direction. Note that $x_f$ has not been changed
for level $2$ even after car $6$ has moved out hence the spacing is remaining same as before.
%
%
%
%
%
%
%
%

\label{sec:conclu}\bibliographystyle{elsarticle-num}

\expandafter\ifx\csname url\endcsname\relax \global\long\def\url#1{\texttt{#1}}
\fi \expandafter\ifx\csname urlprefix\endcsname\relax\global\long\def\urlprefix{URL }
\fi \expandafter\ifx\csname href\endcsname\relax \global\long\def\href#1#2{#2}
 \global\long\def\path#1{#1}
\fi


\end{document}